\documentstyle[epsf]{article}

\title{Competing Species Dynamics: Qualitative Advantage versus
Geography}

\author{
{\bf Serge Galam\footnote{galam@ccr.jussieu.fr}}\\
 Laboratoire des Milieux D\'esordonn\'es et
H\'et\'{e}rog\`enes\thanks{Laboratoire associ\'e au CNRS (UMR
n$^{\circ}$
800) et \`a l'Universit\'e P. et M. Curie - Paris 6},\\ Tour 13 - Case
86,
4 place Jussieu, \\ 75252 Paris Cedex 05, France\\
\\     {\bf  Bastien Chopard\footnote{Bastien.Chopard@cui.unige.ch},}
       {\bf Alexander
Masselot\footnote{Alexandre.Masselot@cui.unige.ch}}\\
D\'epartement d'Informatique, University of Geneva,\\
 24 rue G\'en\'eral-Dufour, 1211 Gen\`eve 4, Switzerland,\\
\\     {\bf Michel Droz\footnote{Michel.Droz@physics.unige.ch}}\\
D\'epartement de Physique Th\'eorique, University of Geneva,\\
 24 quai Ernest-Ansermet, 1211 Gen\`eve 4, Switzerland,}

\date{}

\begin{document}
\maketitle

\begin{abstract}
A simple cellular automata model for a two-group war over the same
``territory'' is presented. It is shown that a qualitative advantage is
not
enough for a minority to win. A spatial organization as well a definite
degree of aggressiveness are instrumental to overcome a less fitted
majority.  The model applies to a large spectrum of competing groups:
smoker-non smoker war, epidemic spreading, opinion formation,
competition
for industrial standards and species evolution. In the last case, it
provides a new explanation for punctuated equilibria.
\end{abstract}

\vskip.5cm

\noindent
Physics has dealt with quite a success in describing and understanding
collective behavior in matter. Very recently many physicists have used
basic concepts and techniques from the physics of collective disorder to
study a large spectrum of problems outside the usual field of physics
such
as social behavior~\cite{kohring:96,glance:93,bonabeau:95}, group
decision making~\cite{galam:97}, financial systems~\cite{levy:95} and
multinational organizations~\cite{galam:96}. See~\cite{stauffer} for a
review of these applications.

A few years ago, Galam has developed a hierarchical voting model based
on
the democratic use of majority rule~\cite{galam:90}. In the simplest
case
of two competing parties $A$ and $B$ with respective support of $a_0$
and
$b_0=1-a_0$, it was shown that, for the $B$, winning the elections at
the
top of the hierarchy (i.e. after several tournaments) does not depend
only
on $b_0$ but also on the existence of some local biases. In particular,
in
the case of voting cells of four persons, a bias is introduced (usually
in
favor of the leading party, e.g. $B$) to solve the $2A$-$2B$ situations.
Then, the critical threshold of support for the ruling party to win can
be
as low as $b_c=0.23$.  The model showed how a majority up to $0.77 $ can
self-eliminate while climbing up the hierarchy, using locally the
democratic majority voting rule. This self-elimination occurs within
only few hierarchical levels.

Following this previous study, we address here the universal and generic
problem of the competing fight between two different groups over a fixed
area.  We present a ``voter model'' which describes the dynamical
behavior of a population with bimodal conflicting interests and
study the conditions of extinction of one of the initial groups.

This model can be thought of as describing the smoker - non smoker
fight:
in a small group of persons, a majority of smokers will usually convince
the few others to smoke and vice versa. The point is really when an
equal
number of smokers and non-smokers meet. In that case, it may be assumed
that a social trend will decide between the two attitudes. In the US,
smoking is viewed as a disadvantage whereas, in France, it is rather
well
accepted. In other words, there is a bias that will select the winner
party in an even situation. In our example, whether one studies the
French
or US case, the bias will be in favor of the
smokers or the non-smokers, respectively.

The same mechanism can be associated with the problem of competing
standards (for instance PC versus Macintosh for computer systems or VHS
versus Beta MAG for video systems). The choice of one or the other
standard
is often driven by the opinion of the majority of people one meets. But,
when the two competing systems are equally represented, the intrinsic
quality of the product will be decisive. Price and technological advance
then play the role of a bias.

Here we consider the case of  four-person confrontations in a spatially
extended system in which the actors (species $A$ or $B$) move randomly.
The process of spatial contamination of opinion plays a crucial role in
this dynamics.

In the original Galam model~\cite{galam:90}, the density threshold for
an
invading emergence of $B$ is $b_c=0.23$ if the $B$ group has a
qualitative bias over $A$. With a spatial distribution of the species,
even if
$b_0<b_c$,  $B$  can still win over $A$ provided that it
strives for confrontation.   Therefore a qualitative
advantage is found not to be enough to win. A geographic as well a
definite
degree of aggressiveness are instrumental to overcome the less fitted
majority.

The model we use to describe the two populations $A$ and $B$ influencing
each other or competing for some unique resources, is based on the
reaction-diffusion automata proposed by Chopard and Droz~\cite{BC-EPL}.
However, here, we consider only one type of particle with two possible
internal states ($\pm1$), coding for the $A$ or $B$ species,
respectively.

The individuals move on a two-dimensional square lattice. At each site,
there are always four individuals (any combination of $A$'s and $B$'s is
possible). These four individuals all travel in a different lattice
direction (north, east, south and west).

The interaction takes place in the form of ``fights'' between the four
individuals meeting on the same site. At each fight, the group nature
($A$
or $B$) is updated according to the majority rule, when possible,
otherwise
with a bias in favor of the best fitted group:
\begin{itemize}
\item The local majority species (if any) wins:
\[ nA+mB\rightarrow \left\{ \begin{array}{ll}
                        (n+m)A & \mbox{if $n>m$} \\
                        (n+m)B & \mbox{if $n<m$} \\
                             \end{array}
\right.
\]
where $n+m=4$.

\item When there is an equal number of $A$ and $B$ on a site, $B$ wins
the confrontation with probability $1/2+\beta/2$. The quantity
$\beta\in[0,1]$ is the bias accounting for some advantage (or extra
fitness) of species $B$.
\end{itemize}
The above rule is applied with probability $k$. Thus, with probability
$1-k$ the group composition does not change because no fight occurs.

Between fights both population agents perform a random walk on the
lattice.
This is achieved by shuffling randomly the directions of motion of the
fours individuals present at each site and letting them move to the
corresponding neighboring sites~\cite{BC-EPL}.

Initially, populations $A$ and $B$ are randomly distributed over the
lattice, with respective concentrations $a_0$ and $b_0=1-a_0$.

It is clear  that the model richness comes from the even
confrontations. If only odd fights would happen, the initial
majority population would always win after some short time.
The key parameters of this model are (i) $k$, the aggressiveness
(probability of confrontation), (ii) $\beta$, the $B$'s bias of winning
a
tie and (iii) $b_0$, the initial density of $B$.

The strategy according to which a minority of $B$'s (with yet a
technical,
genetic, persuasive advantage) can win against a large population of
$A$'s
is not obvious. Should they fight very often, try to spread or accept a
peace agreement? We study the parameter space by running cellular
automata implementing the above system.

In the limit of low aggressiveness ($k\to 0$), the particles move a long
time before fighting. Due to the diffusive motion, correlations between
successive fights are destroyed and $B$ wins provided that $b_0>0.23$
and
$\beta=1.$ This is the mean-field level of our dynamical model which
corresponds to the theoretical calculations made by Galam in his
election
model~\cite{galam:90}.

More generally, and for $\beta={\rm const}$, we observe that $B$ can win
even when $b_0<0.23$, provided it acts aggressively, i.e. by having a
large
enough $k$. Thus, there is a critical density $b_{death}(k)<0.23$ such
that, when $b_0>b_{death}(k)$, all $A$ are eliminated in the final
outcome.
Below $b_{death}$, $B$ looses unless some specific spatial
configurations
of $B$'s are present.

This is a general and important feature of our model: the growth of
species
$B$ at the expense of $A$ is obtained by a spatial organization. Small
clusters that may accidentally form act as nucleus from which the $B$'s
can develop.  In other words, above the mean-field threshold $b_c=0.23$
there is no need to organize in order to win but, below this value only
condensed regions will be able to grow. When $k$ is too small, such an
organization is not possible (it is destroyed by diffusion) and the
strength advantage of $B$ does not lead to success.

\begin{figure}
\centerline{\setlength{\epsfysize}{6cm}
\epsfbox{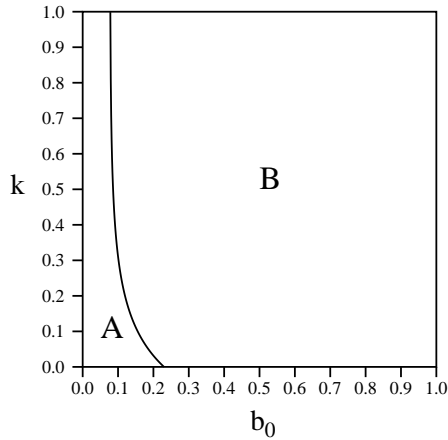}}
\caption{\it  Phase diagram for our socio-physical model with $\beta=1$.
The curve delineates the regions where either $A$ (on the left)
or $B$ (on the right) wins depending on $b_0$, the initial density of
$B$
and $k$, the probability of a confrontation. }
\label{pbvsk0}
\end{figure}

Figure~\ref{pbvsk0} summarizes, as a function of $b_0$ and $k$, the
regions
where either $A$ or $B$ succeeds. It turns out that the separation curve
satisfies the equation $(k+1)^7(b_0-0.077)=0.153$.

It is also interesting to study the time needed to annihilate completely
the looser. Here, time is measured as the number of fights per site
(i.e.
$kt$ where $t$ is the iteration time of the automaton).  We observed
that,
in this case, the dynamics is quite fast and a few units of time are
sufficient to yield a collective change of opinion.

The previous results assume a contant bias. However,
with the assumption that an individual surrounded by several of its
congeners becomes more confident and thus less efficient in its fight,
one may vary the bias $\beta$ as a function of the local density of
$B$.

For example, within a neighborhood of size $\ell^2$, the bias can
decrease
from 1 to 0 as follows : $\beta=1- b/(2\ell^2)$ if $0\le b\le 2\ell^2$
(local minority of $B$'s) and $\beta=0$ if $b>2\ell^2$ (local majority
of
$B$'s), where $b$ designates the number of $B$'s in the neighborhood.

This rule produces an interesting and non-intuitive new behavior.
Depending
on the value of $\ell$, there is a region near $k=1$ such that the $A$
species can win by preventing the $B$'s from spreading in the
environment.
This is achieved by a very aggressive attitude of the $A$'s. Note that
this
effect is already present in the previous case ($\ell=1$ and $\beta={\rm
const}$), but only on the line $k=1$ and for $b_0<0.2$.

\begin{figure}
\centerline{\setlength{\epsfysize}{6cm}
\epsfbox{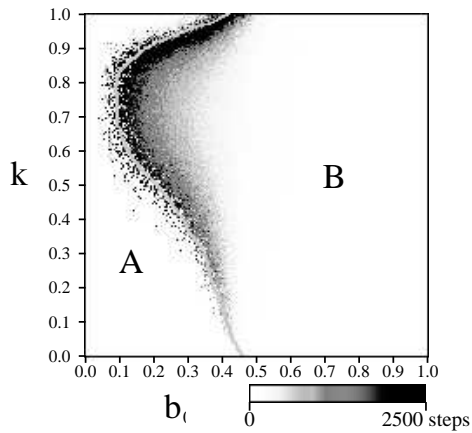}}
\caption{\it Same as figure 1 but for a bias computed
according
to the $B$ density on a local neighborhood of size $\ell=7$. The gray
levels indicate the time to eliminate the defeated species. The black
dots on the left hand side of the separation curve show situation where
the
$B$ species wins due to an accidentally favorable initial configuration
(dark for long time).}
\label{pbvsk3}
\end{figure}

Figure~\ref{pbvsk3} summarizes the regions where either $A$ or $B$
succeeds
when $\ell=7$. In addition to the separation line shown in light gray,
the
time needed to decimate the other opinion is indicated by the gray
levels.
We observe that this time may become large in the vicinity of the
critical
line.  Depending on the time scale associated with the process, such a
slow
evolution may be interpreted as a coexistence of the two species (if a
campaign lasts only a few days or a few weeks, the conflict will not be
resolved within this period of time).

We have shown that the correlations that may exist between
successive fights may strongly affect the global behavior of the
system and that an organization is the key feature to obtain a definite
advantage over the other population. This
observation is important. For instance, during a
campaign against smoking or an attempt to impose a new system, it is
much
more efficient (and cheaper) to target the effort on  small nuclei
of persons rather than sending the information in an uncorrelated
manner.

Also, according to figure~\ref{pbvsk3}, an hypothetical minority of
smokers
in France must harass non-smokers during social meetings (coffee break,
lunch,...) rather often but not systematically, in order to reinforce
their
position. On the contrary, for an hypothetical majority of smokers in
the
US, either a smooth or a stiff harassment against the non-smokers is
required to survive.

Aggressiveness is the key to preserve the spatial organization. Refusing
a
fight is an effective way for the $A$ species to use its numerical
superiority by allowing the $B$ individuals to spread.  With this
respect,
a minority should not accept a peace agreement (which would
results in a lower $k$) with the leading majority unless the strength
equilibrium is modified (i.e. $B$ is better represented).

Motion is also a crucial ingredient in the spreading process. There is
a subtle tradeoff between moving and fighting. When little motion is
allowed between fights ($k\to1$), the advantage is in favor of $A$
again.
In an epidemic system, our model shows that two solutions are possible
to
avoid infestation: either one let the virus die of isolation (dilute
state due to a small $k$) or one decimates it before it spreads (large
$k$).

Finally a simple variant of the above model provides a possible scenario
to
explain punctuated equilibria~\cite{bak-sneppen:93} in the evolution of
living organisms. It is well known that the transition between two forms
of
life may be quite abrupt. There is no trace of the intermediate
evolutionary steps. To give some insights into this problem we modify
our
voter model by including a creation rate for the $B$ individuals ($A\to
B$,
with probability $p\ll 1$).  In this context, the $B$ species is fitter
than
the $A$ species (the bias $\beta=1$) but the numerical advantage of $A$
is
too strong for $B$ to survive. However, if the simulation is run for a
long
enough time, nucleation in this metastable state will happen, which will
produce locally a very favorable spatial arrangement of $B$'s. These
$B$'s
will then develop and, very rapidly, eliminate all $A$'s. In other
words,
a very numerous species may live for a considerable amount of time
without
endangering competitors and suddenly, be decimated by a latent, fitter
species. This scenario needs a strong statistical fluctuation but no
additional external, global event.

In conclusion, although the model we propose is very simple, it
abstracts
the complicated behavior of real life agents by capturing some essential
ingredients. For this reason, the results we have presented may shed
light
on the generic mechanisms observed in a social system of opinion making.

\section*{Acknowledgment} We thank D. Stauffer for a careful reading of
this manuscript. Part of this work was realized during the
``Complexity and Chaos'' workshop at ISI, under grant OFES 95.0046.


\begin{thebibliography}{10}

\bibitem{kohring:96}
G.A. Kohring.
\newblock Ising models of social impact: the role of cumulative
advantage.
\newblock {\em J. Phys. I. France}, 6:301, 1996.

\bibitem{glance:93}
N.S. Glance and B.A. Huberman.
\newblock The outbreak of cooperation.
\newblock {\em J. Math. Sociology}, 17(4):281, 1993.

\bibitem{bonabeau:95}
E.~Bonabeau, G.~Theraulaz, and J.L. Deneubourg.
\newblock Phase diagram of a model of self-organizing hierachies.
\newblock {\em Physica A}, 217:373, 1995.

\bibitem{galam:97}
S.~Galam.
\newblock Rational group decision making: A random field ising model at
$t=0$.
\newblock {\em Physica A}, 238:66--80, 1997.

\bibitem{levy:95}
M.~Levy, H.~Levy, and S.~Solomon.
\newblock Microscopic simulation of the stock market.
\newblock {\em J. de Phys. I (France)}, 5:1087, 1995.

\bibitem{galam:96}
S.~Galam.
\newblock Fragmentation versus stability in bimodal coalitions.
\newblock {\em Physica A}, 230:174--188, 1996.

\bibitem{stauffer}
S.M. de~Oliveira, P.M.C. de~Oliveira, and D.~Stauffer.
\newblock {\em Non-Traditional Applications of Computational Statistical
  Physics: Sex, Money, War, and Computers}.
\newblock Springer, in press.

\bibitem{galam:90}
S.~Galam.
\newblock Social paradoxes of majority rule voting and renormalization
group.
\newblock {\em J. Stat. Phys.}, 61:943--951, 1990.

\bibitem{BC-EPL}
B.~Chopard and M.~Droz.
\newblock Microscopic study of the properties of the reaction front in
an
  {$A+B\rightarrow C$} reaction-diffusion process.
\newblock {\em Europhys. Lett.}, 15:459--464, 1991.

\bibitem{bak-sneppen:93}
P.~Bak and K.~Sneppen.
\newblock Punctuated equilibrium and criticality in a simple model of
  evolution.
\newblock {\em Phys. Rev. Lett.}, 71:4083--4086, 1993.

\end{thebibliography}


\end{document}